\documentclass{jpsj3}
\usepackage{txfonts}
\usepackage{graphicx}
\usepackage{color}

\title{Spatial distribution of reduced density of hard spheres near a hard-sphere dimer: Results from three-dimensional Ornstein-Zernike equations coupled with several different closures and from grand canonical Monte Carlo simulation}

\author{Mika Matsuo$^1$, Yuka Nakamura$^2$, Masahiro Kinoshita$^3$, and Ryo Akiyama$^1$\thanks{rakiyama@chem.kyushu-univ.jp}}
\inst{$^1$Department of Chemistry, Kyushu University, Fukuoka 819-0395, Japan \\
$^2$Department of Physics, Niigata University, Niigata 950-2181, Japan \\
$^3$Department of Chemistry, Graduate School of Science, Chiba University, Chiba 263-8522, Japan \\
Center for the Promotion of Interdisciplinary Education and Research, Kyoto University, Kyoto 606-8501, Japan
} 
\recdate{\today}

\abst{We calculated the spatial distribution of reduced density and pair distribution function (PDF) of solvent hard spheres near a solute using three-dimensional Ornstein-Zernike (OZ) equations coupled with closures in which Percus-Yevick (PY) and hypernetted-chain (HNC) approximations were employed or bridge functions (BFs) proposed by Verlet, Duh and Henderson, and Kinoshita were incorporated. The solute was formed by two solvent hard spheres in contact with each other, with the result that the system is not radial-symmetric, necessitating the extension of Verlet, Duh-Henderson, and Kinoshita BFs to three-variable functions considered in the Cartesian coordinate system. The results were compared with those from grand canonical Monte Carlo (MC) simulation. In terms of the PDF, HNC is superior to PY in the sense that the results from the former are closer to those from MC. The incorporation of the three BFs makes the results further closer to those from MC. The three BFs share almost the same performance for a solute immersed in a one-component solvent at the infinite-dilution limit. Analyses on the triplet distribution function (TDF) were also performed. It was found that in terms of the TDF the incorporation of the BFs does not necessarily lead to improvement.}

\begin{document}
\maketitle
\section{Introduction}
Asakura and Oosawa proposed a mechanism of effective attraction between two large colloidal particles immersed in a polymer solution\cite{asakura1954, asakura1958, vrij1976}. The translational motion of polymers drives the effective attraction. Asakura and Oosawa obtained this effective interaction using the differences in the polymer’s configurational entropy under isochoric condition\cite{footnote1}. Nowadays, this effective interaction is called the depletion interaction\cite{asakura1954, asakura1958, vrij1976, nakamura2019, roth1999, roth2002, rosenfeld1989, minton1997, fulton1982, goodsell199, kinjo2002-1, kinjo2002-2, ellis2003, hall2003, akiyama2006, kinoshita2009, harano2004, roth2006, yoshidome2008, chiba2019, lekkerkerker2011, lekkerkerker1992, frenkel1992, dijkstra1994, dijkstra19942, dijkstra1998, dijkstra19991, dijkstra19992, dijkstra19993, dijkstra2006, cinacchi2007, lopez2013, velasco1999, suematsu2016, kinoshita1994, castaneda2003, roth2005, lekkerkerker2000, lee1999, kubota2012}. On the other hand, a certain configuration of a polyatomic molecule is stabilized arising from the translational motion of the depletants. Which particles are regarded as depletants greatly depends on the respective systems and on the individual researchers' viewpoints. However, the idea of depletion interaction has been applied to explain various phenomena, such as the crowding effects in a living cell\cite{nakamura2019, minton1997, fulton1982, goodsell199, kinjo2002-1, kinjo2002-2, ellis2003, hall2003, akiyama2006}, size separation behaviors\cite{chiba2019, lekkerkerker2011, lekkerkerker1992, frenkel1992, dijkstra1994, dijkstra19942, dijkstra1998, dijkstra19991, dijkstra19992, dijkstra19993, dijkstra2006, cinacchi2007, lopez2013, velasco1999, suematsu2016}, stability of protein conformation\cite{kinoshita2013,harano2005, yoshidome2009, oshima2013, oshima2015, murakami2016, murakami2017}, molecular recognition\cite{kinoshita2013, kinoshita2002, akiyama2010}, and so on.

There is a one-to-one correspondence between the effective interaction and the spatial distribution function, and the spatial distribution functions also influence the dynamical behaviors in a liquid phase. Nakamura et al. showed that the slight difference in the spatial distribution functions around a spherical solute particle strongly affected the diffusion behaviors of the solute\cite{nakamura2021}. 

In the high-density hard-sphere fluid, there is a different feature in the thermodynamic behavior of the large hard-sphere association from that in the dilute fluids. Here, we discuss the association in the hard-sphere fluid, of which the packing fraction is similar to the ambient water. The association in the hard-sphere fluid is entropically driven even under isobaric conditions. On the surface of the large spheres, the density of small hard spheres is relatively higher than that of the bulk. When two large hard spheres contact each other in the fluid under isobaric conditions, some small hard spheres near the surfaces of large hard spheres are released into the bulk. Thus, if the system density increases, the average system volume, namely the partial molar volume of two large hard spheres, does not change in the association process. Some studies indicate that the above story is adequate\cite{yamada2019, inoue2020, baden2007}. We note that, in a fluid consisting of hard-body particle models, such as two large hard-spheres in a small hard-sphere fluid, since all the possible system configurations share the same energy, the system behavior is entropic. On the other hand, calculating the local density, such as the density function near the surface, becomes difficult in the case of high-density fluid.

As the packing fraction of depletants increases, the original Asakura-Oosawa theory worsens\cite{akiyama2006, karino2009}. In the Asakura-Oosawa theory, all the interactions between depletants are ignored in the simple theory. This assumption works in dilute depletant conditions. On the other hand, the deviation of the effective interaction given by the Asakura-Oosawa theory is significant when the depletants are dense like a liquid. In fact, the liquid structure becomes clear with the increase of the packing fraction, and it affects the effective interaction. For example, the liquid structure causes the oscillational structure in the effective interaction\cite{MHNC2002, MHNC2003, akiyama2006, kinoshita2002}. The effective interaction discussed here can be explained by clarifying the spatial distribution of the number density for small spheres in the vicinity of a large sphere or large spheres. Therefore, it is important to calculate them accurately. 

Therefore, we need theoretical methods which predict precise distribution functions around a hard-body solute immersed in a hard sphere fluid in the context of the depletion interaction. This is because the repulsive part of the direct interaction has an essential role in forming a microscopic liquid structure \cite{chandler1970, weeks1971, chandler1983}. In this respect, a hard-sphere system serves as the most fundamental model in the studies of a dense fluid. Although the precise prediction becomes difficult when the packing fraction is high, like a liquid, there are some precise theories, such as density functional theories (DFT)\cite{rosenfeld1989, roth2002, yu2002, roth1999} and integral equation theories\cite{hansen1986, attard1990, kinoshita1996}.

The present study carries out to discusses the accuracy of the solvent density profiles around a contact dimer calculated using some integral equation theories. The Ornstein-Zernike (OZ) equation is solved with a closure relation\cite{hansen1986}, and the accuracy depends on the closure relation. It is known that some traditional closures, such as the Percus-Yevick (PY) closure and the hypernetted-chain (HNC) closure, give us qualitatively adequate spatial distribution functions. However, we can find differences between them. 
It is known that the PY closure is better than the HNC closure for the one-component hard-sphere fluid. On the other hand, the HNC closure has been adopted in the calculations of hard-sphere mixtures because it is better than the PY closure when the size ratio is not 1. We also examined the MHNC closure proposed by Kinoshita in our previous study\cite{MHNC2002, MHNC2003, kinoshita2017, nakamura2019}. In the study, a hard-sphere solute was immersed in a hard-sphere fluid of which the packing fraction was 0.38, and various size solutes were examined. The results show that the spatial distribution function calculated by the MHNC-OZ theory is much better than that calculated by the PY-OZ and the HNC-OZ theories. Moreover, the MHNC closure is accurate even when the size asymmetry is significant.

In the present study, we examine non-spherical solutes, namely a contact dimer of solvent particles. As we mentioned above, MHNC closure is an accurate closure for a spherical solute in a solvent hard-sphere fluid. However, study on the accuracy of distribution functions in the vicinity of a non-spherical solute is unusual because the calculation cost of molecular simulation is very expensive, and we must obtain the spatial distribution function in the vicinity of a non-spherical solute for the coordinate $(x,y,z)$ in the calculation using the integral equation theory. If we take into account symmetry, we can reduce the calculation cost. In this study, we carried out the simulation to compare the result obtained using the three-dimensional (3D) integral equation theory. We carry out Grand Canonical Ensemble Monte Carlo simulations\cite{frenkel1992, adams1974chemical, adams1975grand} to verify the 3D-MHNC-OZ theory in the present study.

\section{Model and Methods}
\subsection{Model}
A fluid is composed of hard-spheres. The diameter was $\sigma_V$ ($V$ denotes the solvent particles). The scaled number density of the fluid ${\rho \sigma_V}^3$ was at 0.7315. This density of the fluid corresponds to the packing fraction $\eta=0.383$. This value is the packing fraction of pure water at standard temperature and pressure. We immersed a solute into the hard-sphere fluid. The solute was a contact dimer which consisted of two hard-spheres fixed at  coordinates $(-0.5\sigma_V,0,0)$ and $(0.5\sigma_V,0,0)$. Each diameter of  two hard-spheres was $\sigma_U=\sigma_V$ ($U$ denotes solute particles).

\subsection{Integral Equation Theory}
In the present study, we solved the OZ equation with a closure equation to obtain the spatial distribution $g_{UV}(x,y,z)$ of solvent particles around a solute. The OZ equation for the bulk solvent was
\begin{equation}
\label{OZ}
    h_{VV}(r)=c_{VV}(r)+ \rho_V \int c_{VV}(r)h_{VV}(|\mathbf{r}'-\mathbf{r}|) d\mathbf{r}'
\end{equation}
where $\rho$ is the number density, $h$ is the total correlation function, $c$ is the direct correlation function. The system is spherically symmetric and $r$ is the distance between centers of solvent particles. At first, we solved this equation with a closure equation.

The solute particle has a non-spherical shape. Therefore, the spatial distribution functions between a solute and solvent particles are $g_{UV}(x, y, z)= h_{UV}(x, y, z)+1$. To solve the equations, we prepared a 3D grid covering enough volume. Then, the OZ equation is given in the discrete form as follows,
\begin{equation}
\label{3D-OZ}
    h_{UV}(x,y,z)=c_{UV}(x,y,z)+ \rho_V \int c_{UV}(x,y,z)h_{VV}(|\mathbf{r}'-\mathbf{r}|)dx'dy'dz'
\end{equation}
where ${\mathbf{r}}=(x,y,z)$ and ${\mathbf{r'}}=(x',y',z')$. Eq.(\ref{3D-OZ}) is written in the wavenumber $\mathbf{k}$ space as follows:
\begin{equation}
\label{fourier}
\hat{\gamma}_{UV}(k_x,k_y,k_z)=\rho_V \hat{c}_{UV}(k_x,k_y,k_z)\hat{h}_{VV}(|\mathbf{k}|)
\end{equation}
where $\gamma=h-c$, the symbol "\textasciicircum" indicates the Fourier transform. The vector $\mathbf{k}=(k_x,k_y,k_z)$.

We examined some closures. The closure equation is written as
\begin{equation}
\label{closure}
    c_{ij}(\mathbf{r})=\exp[-\beta u_{ij}(\mathbf{r})]\exp[\gamma_{ij}(\mathbf{r})+b_{ij}(\mathbf{r})]-\gamma_{ij}(\mathbf{r})-1
\end{equation}
where $\beta=(k_BT)^{-1}$, $k_BT$ is Boltzmann constant times the absolute temperature. The functions $u$ and $b$ are the potential and the bridge function, respectively. In the calculation of bulk solvent, $i=j=V$ and $\mathbf{r}$ is replaced with $r$. On the other hand, $i=U$, $j=V$ and ${\mathbf{r}} = (x, y, z)$ in the calculation of the solute-solvent correlation functions.

Eq.(\ref{closure}) includes the bridge function. If we had perfect bridge function, 
the closure would be perfect too. However, the exact form of the bridge function has not been known yet. In the case of the HNC approximation,
\begin{equation}
    b_{ij}(\mathbf{r})=0.
    \label{HNCeq}
\end{equation}
In the present study, we examine the MHNC closure proposed by Kinoshita\cite{MHNC2002, MHNC2003, nakamura2019}. The bridge function is,
\begin{equation}
\label{MHNC}
    \begin{split}
        b_{ij}(\mathbf{r})&=-0.5\frac{\gamma_{ij}^2(\mathbf{r})}{1+0.8\gamma_{ij}(\mathbf{r})}(\gamma_{ij}>0)\\
        &=-0.5\frac{\gamma_{ij}^2(\mathbf{r})}{1-0.8\gamma_{ij}(\mathbf{r})}(\gamma_{ij}<0).
    \end{split}
\end{equation}

We also examine following two bridge functions. The bridge function proposed by Verlet is as follows\cite{verlet}:
\begin{equation}
\label{Verlet}
        b_{ij}(\mathbf{r})=-0.5\frac{\gamma_{ij}^2(\mathbf{r})}{1+0.8\gamma_{ij}(\mathbf{r})}.
\end{equation}
Another bridge function proposed by Duh and Henderson is as follows\cite{duh}:
\begin{equation}
\label{DH}
    \begin{split}
        b_{ij}(\mathbf{r})&=-0.5\frac{\gamma_{ij}^2(\mathbf{r})}{1+0.8\gamma_{ij}(\mathbf{r})}(\gamma_{ij}>0)\\
        &=-0.5\gamma_{ij}^2(\mathbf{r})(\gamma_{ij}<0).
    \end{split}
\end{equation}

We also examine the PY closure:
\begin{equation}
c_{ij}(\mathbf{r})=\exp[-\beta u_{ij}(\mathbf{r})][\gamma_{ij}(\mathbf{r})+1]-\gamma_{ij}(\mathbf{r})-1.
\end{equation}
It is known that the PY closure is excellent for the one-component hard-sphere fluid.
The spatial distribution of solvents $g_{UV}(x,y,z)$ is obtained using calculated $\gamma_{UV}(x, y, z)$ and $c_{UV}(x, y, z)$ as follows:
\begin{equation}
\label{sdf}
    g_{UV}(x,y,z)=\gamma_{UV}(x,y,z)+c_{UV}(x,y,z)+1.
\end{equation}

The numerical procedures are as follows: (a) $h_{VV}(r)$ is calculated using Eq.(\ref{OZ}) and one of the closures, (b) $h_{VV}(x, y, z)$ is prepared using $h_{VV}(r)$ and transformed to $\hat{h}_{VV}(k_x, k_y, k_z)$ using the 3D fast Fourier transform (3D-FFT), (c) $u_{UV}(x, y, z)$ is calculated at each 3D grid points and $\gamma_{UV}(x,y,z)$ is initialized to zero, (d) $c_{UV}(x, y, z)$ is calculated using Eq.(\ref{3D-OZ}) and the same closure adopted in step (a), (e) $c_{UV}(x, y, z)$ is transformed to $\hat{c}_{UV}(x, y, z)$ using the 3D-FFT, (f) $\hat{\gamma}_{UV}(x,y,z)$ is calculated using Eq.(\ref{fourier}), (g) $\hat{\gamma}_{UV}(x,y,z)$ is transformed to $\gamma_{UV}(x, y, z)$ using the inversed 3D-FFT and steps (d)-(g) are repeated until the difference between the input and output functions become smaller than the given value, (h) $g_{UV}(x,y,z)$ are calculated using Eq.(\ref{sdf}).
The grid spacing $(\Delta x,\Delta y,\Delta z)$ is $0.02\sigma_V$ and the grid resolution $(N_{x},N_{y},N_{z})$ is $512$.

\subsection{MC simulation}
We fixed the basic cell size and adopted the periodic boundary condition. In the present study, three types of MC simulations were carried out. We examined the canonical MC (CMC) simulations and the grand canonical MC (GCMC) simulation. The canonical MC has a problem in comparison with the results calculated by the integral equations. The integral equation theory is formulated using the grand canonical ensemble, and the solvent density $\rho_V$ is determined at the reservoir. On the other hand, the solvent number density in the basic cell for the canonical MC deviates due to the insertion of the solute particle into the fluid. Then, we must obtain the number of solvent particles and the volume of the basic cell. However, it is not easy in general. Fortunately, Schmidt and Skinner proposed a recipe\cite{schmidt2003hydrodynamic}. Therefore, we adopted the recipe and adjusted the cell volume by using the following rule,
\begin{equation}
    V= \frac{N}{\rho_V}+\Delta V_{ex}
\end{equation}
where $N$ is total number (solvent and solute) of particles and $\Delta V_{ex}$ is the difference between the excluded volumes of solute and solvent particle.
If the solute particle is spherical, $\Delta V = \frac{\pi}{6}[(\sigma_U+\sigma_V)^3 - (\sigma_V+\sigma_V)^3]$, where $\sigma_U$ and $\sigma_V$ are the solute and the solvent diameters, respectively. In the case of a spherical solute particle, this recipe has been adopted, and it has given us satisfactory results\cite{schmidt2003hydrodynamic, sokolovskii2006tracer, nakamura2019}. We adopted this recipe, although the solute shape was not spherical in the present study. 


We also carried out the grand canonical MC (GCMC)\cite{frenkel1992, adams1974chemical, adams1975grand, nezbeda1991}. Although the calculation cost of the GCMC is relatively expensive, this choice is most adequate because the integral equation theory is formulated using the grand canonical ensemble. 

We consider a one-component system with fixed volume($V$), temperature($T$), and chemical potential($\mu$). $\mathbf{x}_i$ denotes a given configuration $i$ of the system, and $U(\mathbf{x}_i)$ is the corresponding total potential energy. The probability $P_i$ of configuration $\mathbf{x}_i$ in the grand canonical ensemble is given by

\begin{equation}
    P_i=\frac{1}{\Xi}\frac{1}{N!\Lambda^{3N}}\exp[\beta N\mu-\beta U(\mathbf{x}_i)].
\end{equation}
$\Xi$ is the grand canonical partition function
\begin{equation}
    \Xi =\sum_{N=0}^{\infty} \frac{1}{N!\Lambda^{3N}} \int \ldots \int \exp[\beta N\mu-\beta U(\mathbf{x}_i)]d\mathbf{x}_1 \ldots d\mathbf{x}_N
\end{equation}
where $\Lambda=h/(2\pi m k T)^{1/2}$. $h$ and $m$ are Plank's constant and the mass of a particle, respectively. The chemical potential for an ideal gas is
\begin{equation}
    \mu_{id}=kT[\ln{\Lambda}^3+\ln{\frac{\bar{N}}{V}}]
\end{equation}
where $\bar{N}$ is the average number of particles. Transforming the integrals to dimensionless particle coordinates, $d\mathbf{\tau}=V^{-1}d\mathbf{x}$, and substituting the chemical potential for an ideal gas produces
\begin{equation}
\label{prob}
    P_i=\frac{1}{\Xi}\frac{1}{N!}\exp[\beta N\mu_{ex}-\beta U(\mathbf{\tau}_i)+N\ln{\bar{N}}].
\end{equation}
In Eq.(\ref{prob}), $\mu_{ex}$ is the excess chemical potential of fluid relative to a perfect gas with the same particle mass, density and temperature. Before GCMC simulation, we used Widom's insertion method in a canonical ensemble for determining the excess chemical potential as a function of the bulk density.

\begin{figure}[h]
 \begin{minipage}[b]{0.48\columnwidth}
  \centering
  \includegraphics[width=\columnwidth]
  {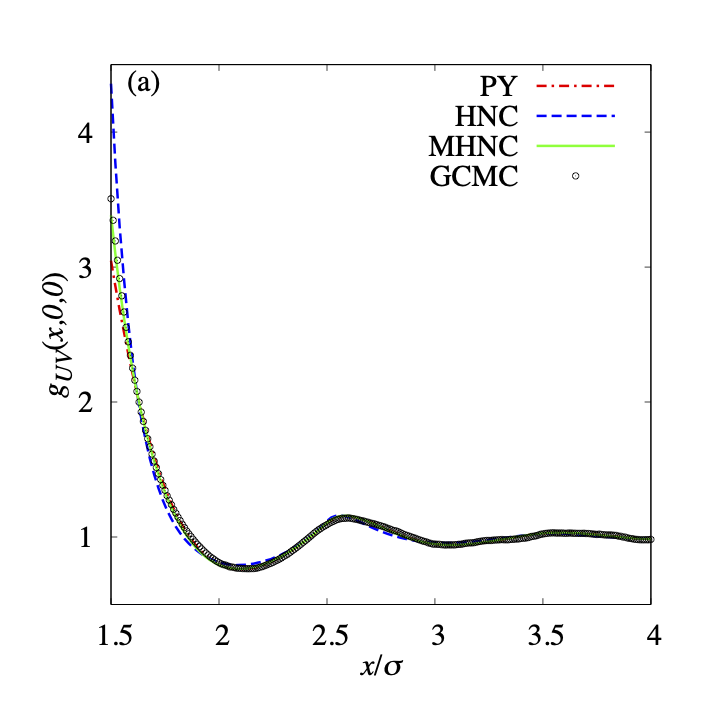}
 \end{minipage}
 \begin{minipage}[b]{0.5\columnwidth}
  \centering
  \includegraphics[width=\columnwidth]
  {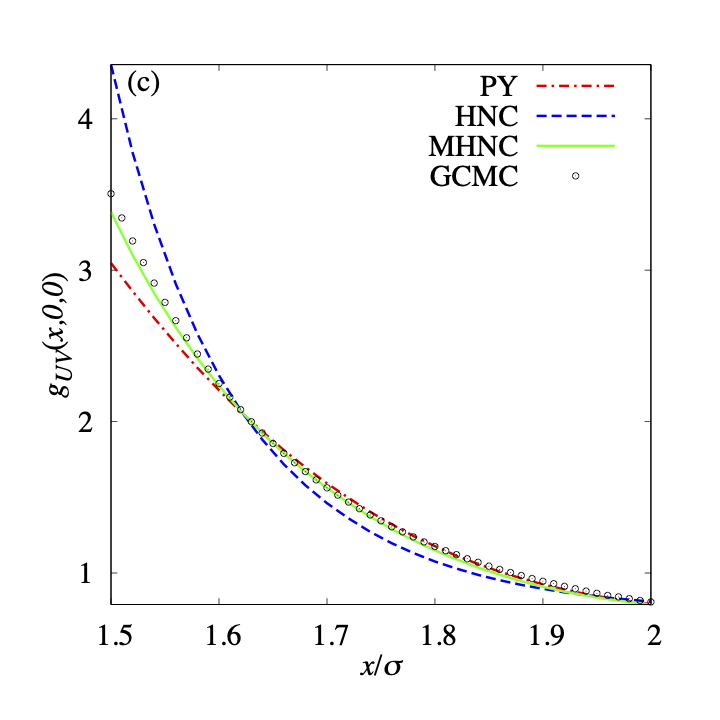}
 \end{minipage}\\
 \begin{minipage}[b]{0.48\columnwidth}
  \centering
  \includegraphics[width=\columnwidth]
  {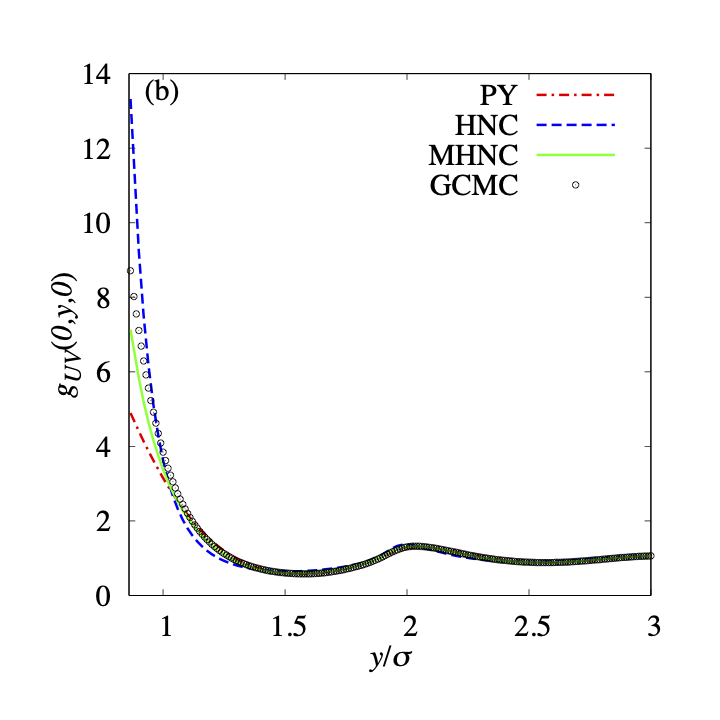}
 \end{minipage}
 \begin{minipage}[b]{0.48\columnwidth}
  \centering
  \includegraphics[width=\columnwidth]
  {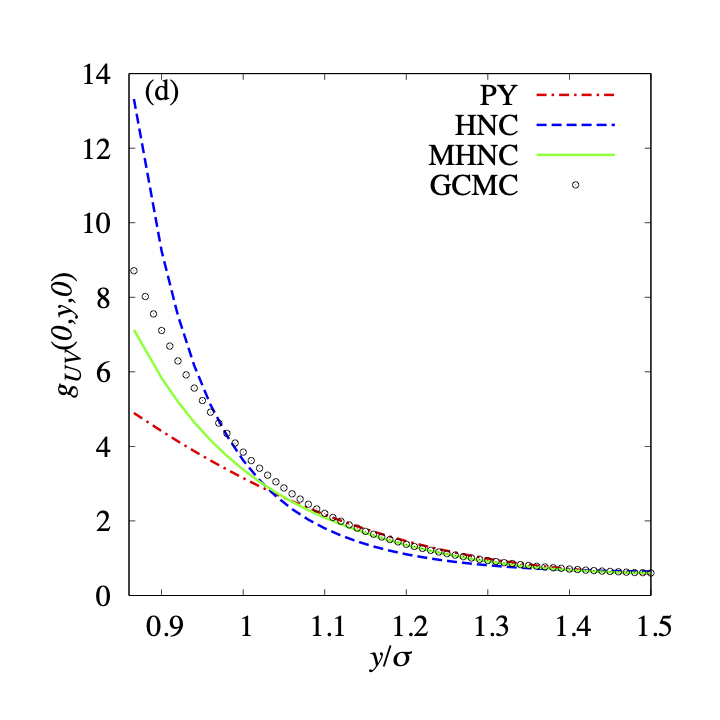}
 \end{minipage}
 \caption{The spatial distribution (a) along the $x$-axis $g_{UV}(x,0,0)$ and that (b) along the $y$-axis $g_{UV}(0,y,0)$. The results of CMC and GCMC are compared. (c)$-$(d) are magnifications near the contact of (a)$-$(b), respectively.}
 \label{mcgxy}
\end{figure}
The grid cell volume is $\Delta V=\Delta x\Delta y\Delta z$ where $\Delta x$, $\Delta y$, and $\Delta z$ denote the grid spacings in the Cartesian coodinate and $\Delta x=\Delta y=\Delta z= 0.01\sigma_V$.  The cell size was set equal to $L=12.36\sigma_V$ and the number of particles $N$ was $1372$ in the CMC simulation. The number of particles varies in the GCMC simulation.
The spatial distribution function $g(x,y,z)$ was calculated as follows:
\begin{equation}
g_{US}(x,y,z)=\Delta N(x,y,z)/\rho \Delta V
\end{equation}

where $\Delta N(x,y,z)$ is number of solvent particles in the grid cell. We used the Metropolis algorithm and performed $10^{7}$ MC steps for equilibration of hard-sphere fluids and over $10^{10}$ MC steps for collection of ensemble averages. The contact value was estimated by extrapolation\cite{nakamura2019}.

We compared the distribution functions around the dimer obtained using CMC and GCMC in Fig.\ref{mcgxy} (for another CMC results, see Fig.S1 of the Supporting Information). The agreement was good. If the sampling was hard, we could use CMC. Since the results of CMCs and GCMC agreed well as mentioned above, we can conclude that the size of the simulation box is large enough and the sampling number is sufficiently enough to obtain the distribution functions. Then, the GCMC results can be recognized as exact results.
\section{Results and Discussion}\label{sec3}
\subsection{Calculation for a spherical solute}

\begin{figure}[h]
    \centering
    \includegraphics[width=7.5cm]{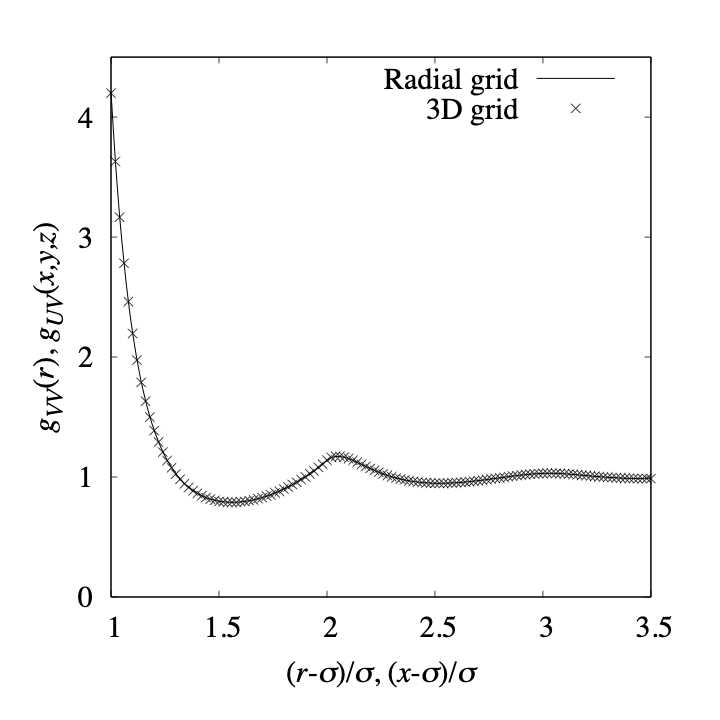}
    \caption{The distribution function of solvents around a spherical solute obtained from the integral equation theory on a radial grid with spacing 0.01$\sigma_V$(solid line) and 3D grid with spacing 0.02$\sigma_V$(crocces). The HNC closure is used.}
    \label{1s}
\end{figure}

To test the feasibility of the numerical solution of integral equation theory on a three-dimensional discretized grid, the solvent distribution around a single particle is considered. Since the grid points are finite, we examine the numerical precision. If the solute is spherical, the results of the three-dimensional integral equation theory and the radial-symmetric integral equation theory must agree. The distribution function of solvents along $x$-axis through the center of the particle is shown in Fig.\ref{1s}. The plot agrees with the result calculated using the radial-symmetric integral equation theory. We confirmed that the 3D integral equation theory accurately reproduced the correct density distribution.

\subsection{Spatial distribution of solvent around a contact dimer}

\begin{figure}[h]
 \begin{minipage}[b]{0.48\columnwidth}
  \centering
  \includegraphics[width=\columnwidth]
  {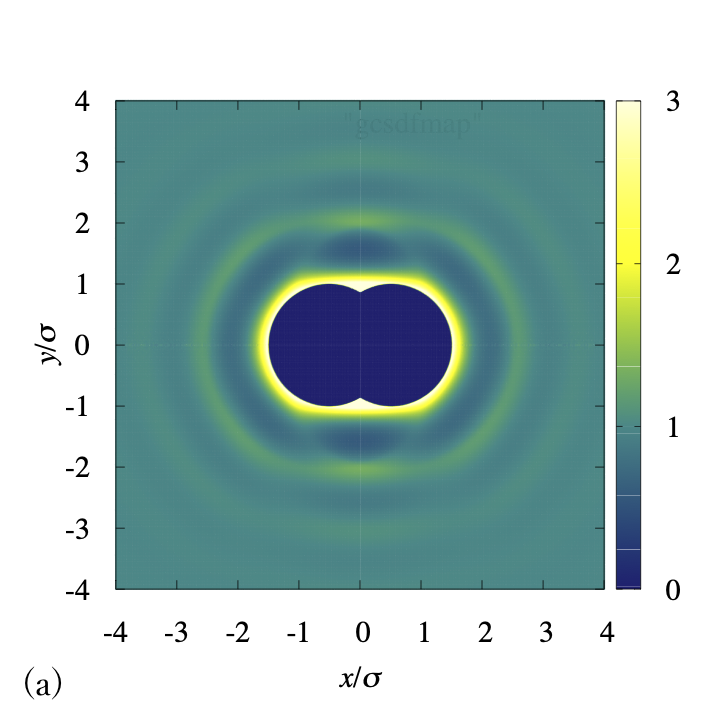}
 \end{minipage}
 \begin{minipage}[b]{0.5\columnwidth}
  \centering
  \includegraphics[width=\columnwidth]
  {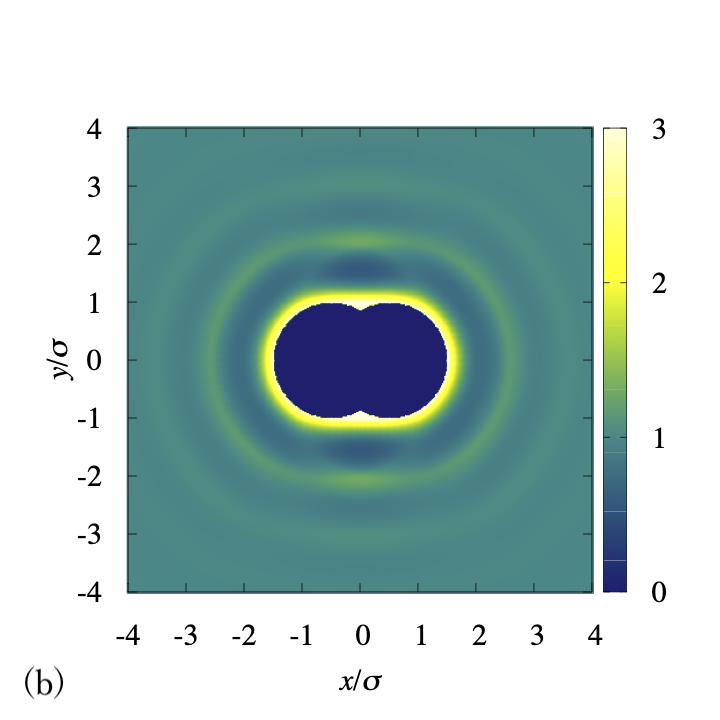}
 \end{minipage}
 \caption{The spatial distribution around dimer in the $xy$ plane $g_{UV}(x,y,0)$ calculated by (a)  GCMC simulation and (b) PY closure.}
 \label{map}
\end{figure}

We obtained the spatial distribution $g_{UV}(x,y,z)$ around non-spherical solute, a contact dimer, calculated using GCMC simulation. Fig.\ref{map}(a) shows the spatial distribution around the dimer $g_{UV}(x,y,0)$. This distribution function is compared with those obtained using the 3D integral equation theories. The result obtained using PY closure is shown in Fig.\ref{map}(b). This color map for the solvent distribution is similar to each other. This agreement means that PY closure is reasonable qualitatively. The spatial distributions $g_{UV}(x,y,0)$ calculated using other closures are also similar.

\begin{figure}[h]
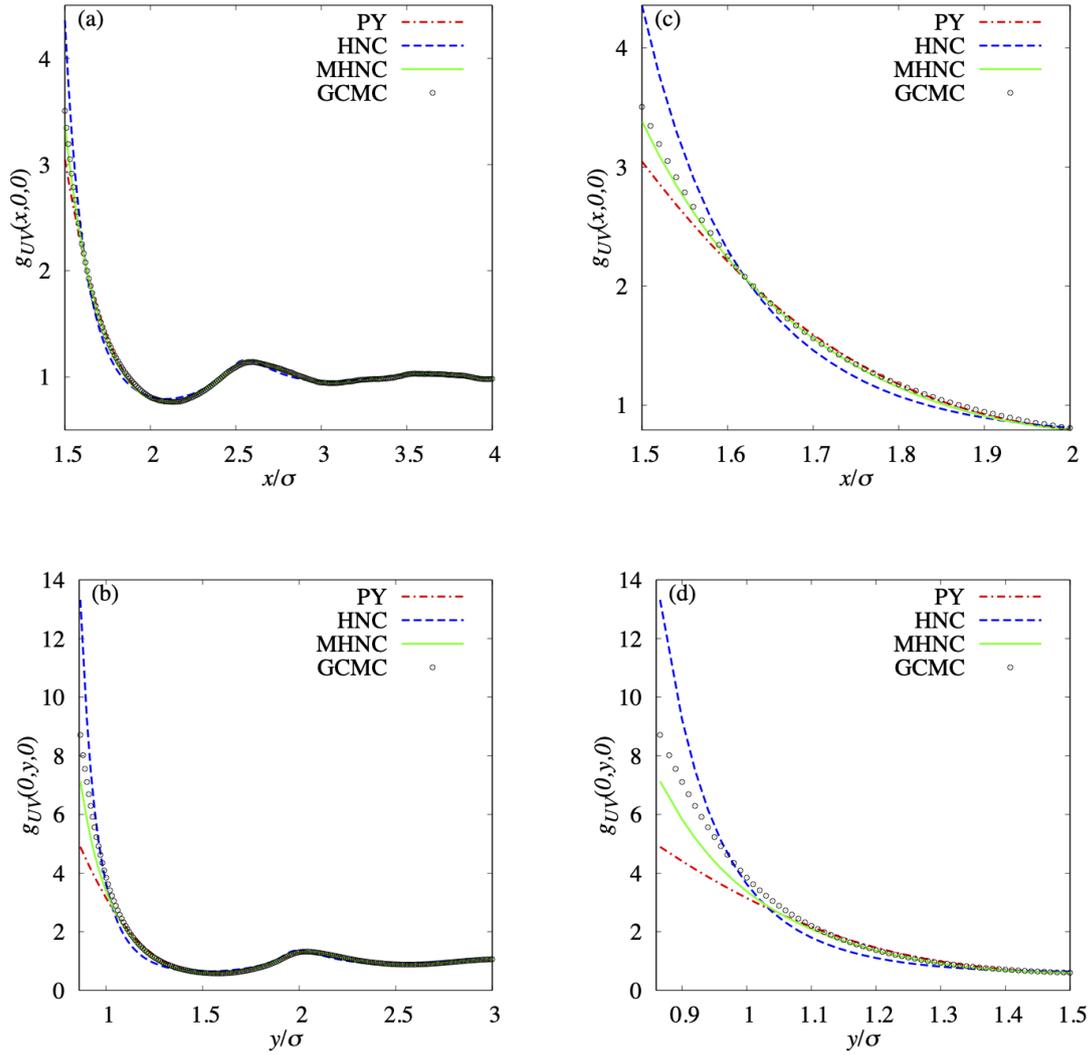

 \begin{minipage}[b]{0.48\columnwidth}
  \centering
  \includegraphics[width=\columnwidth]
  {gx.png}
 \end{minipage}
 \begin{minipage}[b]{0.48\columnwidth}
  \centering
  \includegraphics[width=\columnwidth]
  {gxc.png}
 \end{minipage}\\
 \begin{minipage}[b]{0.48\columnwidth}
  \centering
  \includegraphics[width=\columnwidth]
  {gy.png}
 \end{minipage}
 \begin{minipage}[b]{0.48\columnwidth}
  \centering
  \includegraphics[width=\columnwidth]
  {gyc.png}
 \end{minipage}
 \caption{The spatial distribution (a) along the $x$-axis $g_{UV}(x,0,0)$ and that (b) along the $y$-axis $g_{UV}(0,y,0)$. The results of integral equation theory with the PY, HNC and MHNC closures are compared with the GCMC simulation. (c)$-$(d) are magnifications near the contact of (a)$-$(b), respectively.}
 \label{gxy}
\end{figure}

To quantitatively compare the results obtained by MC and closures we plot the spatial distribution along the $x$-axis $g_{UV}(x,0,0)$ and that along the $y$-axis $g_{UV}(0,y,0)$ in Fig.\ref{gxy}. We chose two specific sections, namely most convex and concave sections, and plotted the distribution functions, namely $g_{UV}(x,0,0)$ and $g_{UV}(0,y,0)$.

Fig.\ref{gxy}(a), (c) shows the solvent distribution around the most convex surface of the solute, namely $g_{UV}(x,0,0)$. The curvature of the solute particle is the same as that of the solvent particles. We can find the difference near the contact distance. The plot calculated by MHNC approximation almost agrees with that calculated by GCMC simulation.On the other hand, the HNC approximation overestimates, and the PY approximation underestimates the values near the solute surface. This agreement and these differences were also observed when the solute is a spherical one that has solvent particle size\cite{nakamura2019}. These results are reasonable because the curvature of the most convex surface of the solute is the same as that of the solvent particles. 

On the other hand, the behaviors of $g(0,y,0)$ (Fig.\ref{gxy}(b),(d)) deviate from those of $g(x,0,0)$. The differences from the GCMC simulation result become significant. At the contact distance, the value of the HNC approximation is about $1.5$ times, and the value of the PY approximation is about half of the GCMC result. Only the MHNC result keeps a small deviation from that for GCMC. The ratio of the contact value is about $0.85$. It indicates that MHNC closure gives a good approximation near the concave surface.

\begin{figure}[h]
 \begin{minipage}[b]{0.48\columnwidth}
  \centering
  \includegraphics[width=\columnwidth]
  {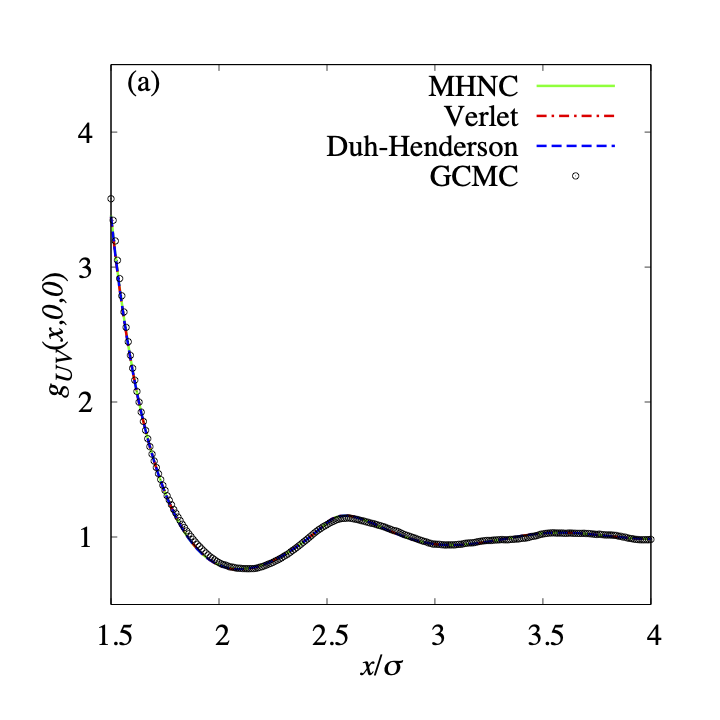}
 \end{minipage}
 \begin{minipage}[b]{0.48\columnwidth}
  \centering
  \includegraphics[width=\columnwidth]
  {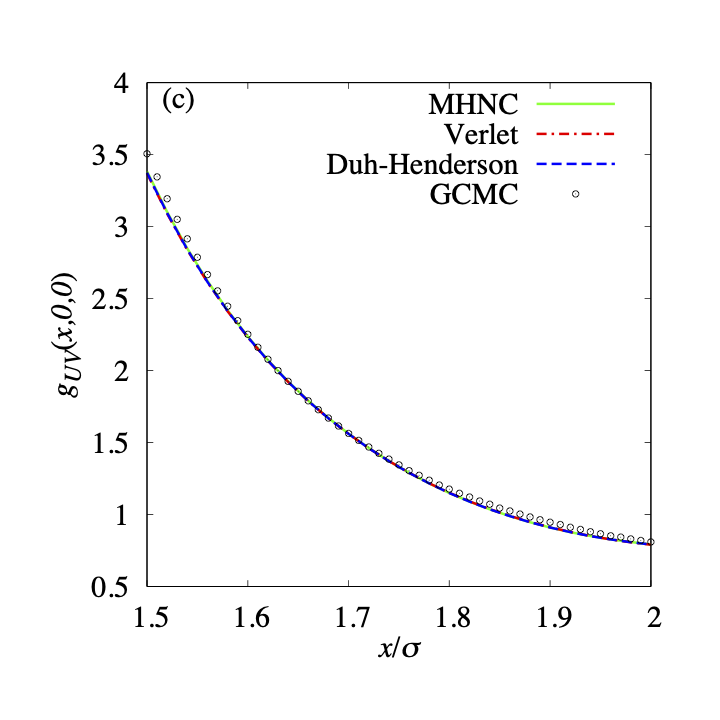}
 \end{minipage}\\
 \begin{minipage}[b]{0.48\columnwidth}
  \centering
  \includegraphics[width=\columnwidth]
  {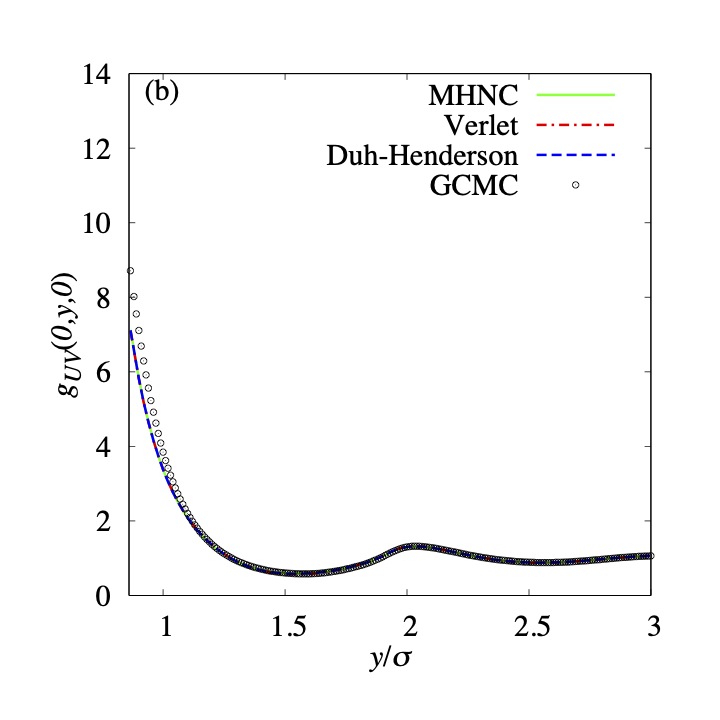}
 \end{minipage}
 \begin{minipage}[b]{0.48\columnwidth}
  \centering
  \includegraphics[width=\columnwidth]
  {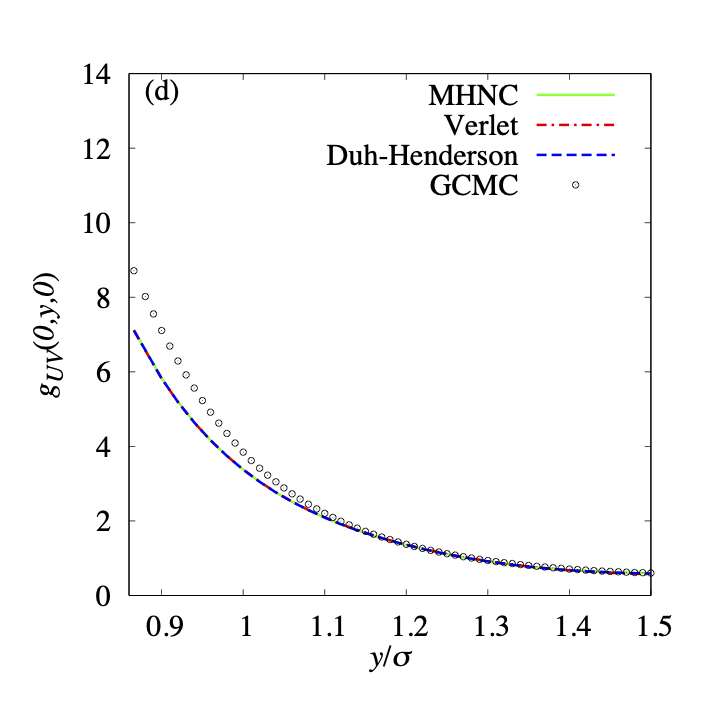}
 \end{minipage}
 \caption{The spatial distribution (a) along the $x$-axis $g_{UV}(x,0,0)$ and that (b) along the $y$-axis $g_{UV}(0,y,0)$. The results of integral equation theory with the MHNC, Verlet and Duh-Henderson bridge functions are compared with the GCMC simulation. (c)$-$(d) are magnifications near the contact of (a)$-$(b), respectively.}
 \label{gxy-VD}
\end{figure}

We also examine other bridge functions, namely, the Verlet bridge function and the Duh-Henderson bridge function. Fig.\ref{gxy-VD} shows the results. We cannot distinguish the plots for the three bridge functions (Eqs.\ref{MHNC}, \ref{Verlet}, \ref{DH}). These bridge functions were also examined in the previous papers\cite{MHNC2003, nakamura2019}. When a spherical solute is immersed in the one-component hard sphere fluid, these bridge functions gave us adequate spatial distribution functions. Therefore, we discuss that the superiority of the MHNC closure is slight in the three closures.  The superiority of the MHNC closure becomes significant when the value of $\gamma_{ij}(r)$ is negative and the absolute value  $|\gamma_{ij}(r)|$ is large enough. As we mentioned in the previous paper \cite{nakamura2019}, the function  $\gamma_{ij}(r)$ does not have a large negative value in the case of a one-component solvent system. This is the reason for the small differences between the results calculated using three bridge functions. In the present study, this conclusion is maintained even when the surface of the solute is concave in the case of $g(0,y,0)$ (See Fig.\ref{gxy-VD}(b), (d)). The accuracies of the approximation with the bridge functions depend on the surface curvature of the solute. Fig.\ref{gxy} suggests that the accuracy near the convex surface of the solute is better than that near the concave surface. The spatial distribution functions $g(0,y,0)$ obtained using the approximation with the bridge functions are slightly smaller than that obtained using GCMC.

\subsection{Triplet distribution functions}

\begin{figure}[h]
    \centering
    \includegraphics[width=7.5cm]{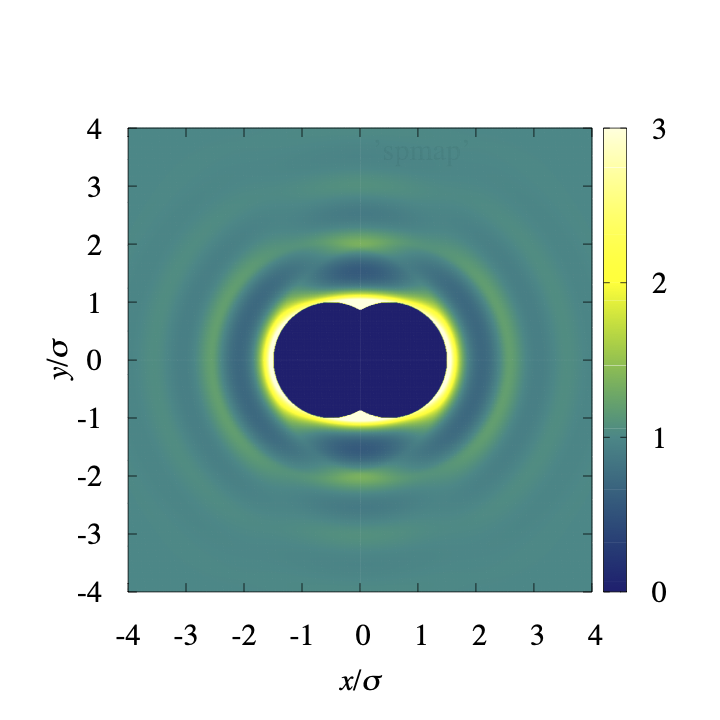}
    \caption{The spatial distribution obtained by using the superposition approximation (SA) around dimer in the $xy$ plane $g(x,y,0)$.}
    \label{spmap}
\end{figure}

\begin{figure}[h]
 \begin{minipage}[t]{0.48\columnwidth}
  \centering
  \includegraphics[width=\columnwidth]
  {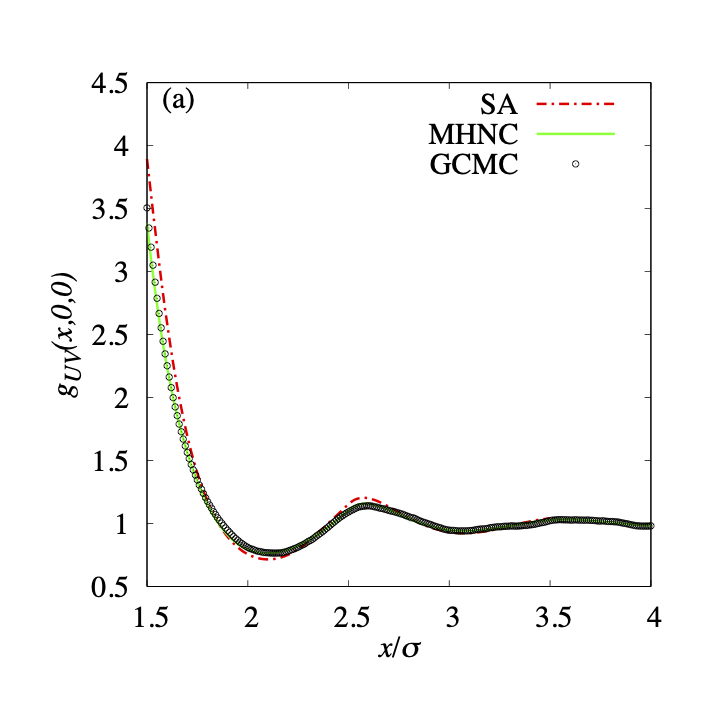}
 \end{minipage}
 \begin{minipage}[t]{0.48\columnwidth}
  \centering
  \includegraphics[width=\columnwidth]
  {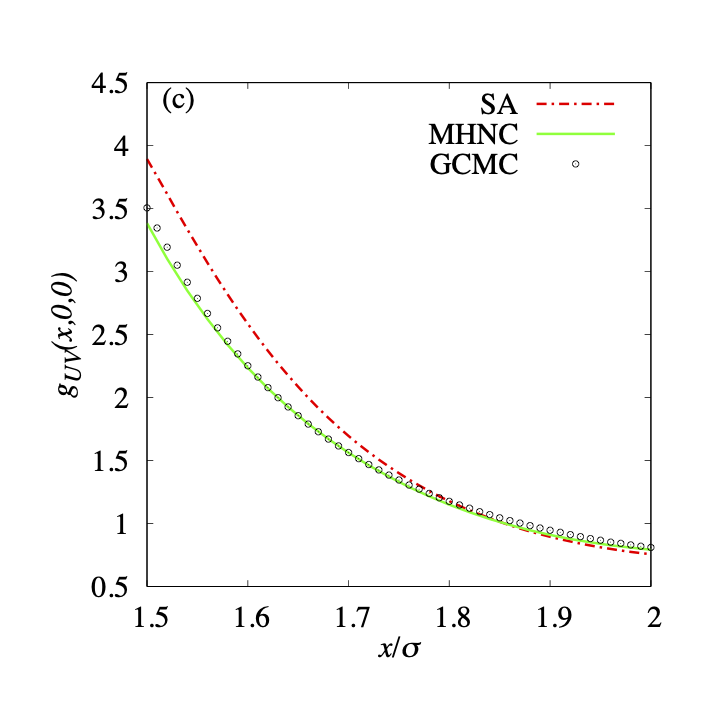}
 \end{minipage}\\
 \begin{minipage}[t]{0.48\columnwidth}
  \centering
  \includegraphics[width=\columnwidth]
  {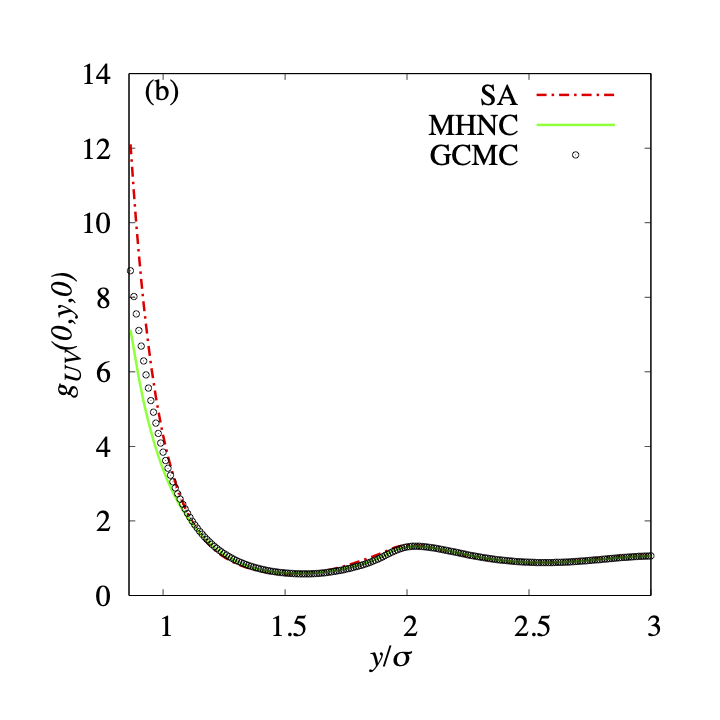}
 \end{minipage}
 \begin{minipage}[t]{0.48\columnwidth}
  \centering
  \includegraphics[width=\columnwidth]
  {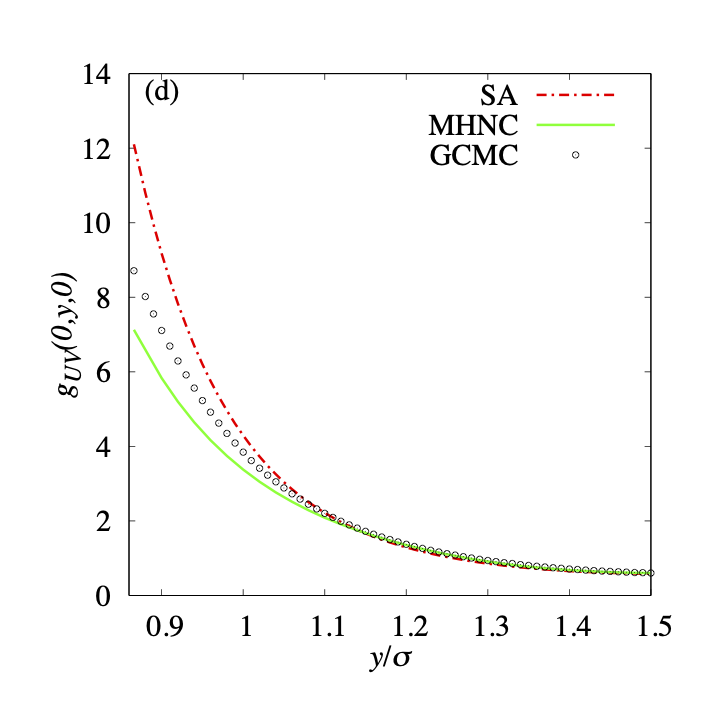}
 \end{minipage}
 \caption{The spatial distribution along (a) the $x$-axis $g_{UV}(x,0,0)$ and that along (b) the $y$-axis $g_{UV}(0,y,0)$. (c)$-$(d) are magnifications near the contact of (a)$-$(b), respectively. Here, $g(r)$ obtained by GCMC is adopted in SA calculation.}
 \label{gxy_SA}
\end{figure}

The validity of the closure can also be evaluated in terms of the three-body correlation. We discuss the triplet distribution functions. The triplet distribution functions $g^{(3)}(\mathbf{r_1},\mathbf{r_2},\mathbf{r_3})$ mean the reduced probability of finding three particles at positions $\mathbf{r_1}$, $\mathbf{r_2}$, and $\mathbf{r_3}$\cite{alder1964triplet}. Then $g^{(3)}$ can be expressed by a triple product of the pair distribution functions $g^{(2)}$ which is called superposition approximation (SA)\cite{kirkwood1935statistical};

\begin{equation}
\label{g3}
    g^{(3)}_{\rm{SA}}(\mathbf{r_1},\mathbf{r_2},\mathbf{r_3})=g(\mathbf{r_1},\mathbf{r_2})g(\mathbf{r_2},\mathbf{r_3})g(\mathbf{r_3},\mathbf{r_1}).
\end{equation}
When two particles are fixed at $\mathbf{r_1}$ and $\mathbf{r_2}$, reduced probability of finding a particle at position $\mathbf{r_3}$ is as follows using Eq.(\ref{g3});

\begin{equation}
\label{SA}
    \begin{split}
        g_{\rm{SA}}(r_{12};\mathbf{r_3})&=g^{(3)}_{\rm{SA}}(\mathbf{r_1},\mathbf{r_2},\mathbf{r_3})/g(\mathbf{r_1},\mathbf{r_2})\\
        &=g(\mathbf{r_2},\mathbf{r_3})g(\mathbf{r_3},\mathbf{r_1}) 
    \end{split}
\end{equation}
%
where
\begin{equation}
    r_{12}\equiv |\mathbf{r_1}-\mathbf{r_2}|.
\end{equation}
Choosing the third Cartesian coordinate $\mathbf{r_3}=(x_3,y_3,z_3)$ in Eq.(\ref{SA}) and setting $\mathbf{r_1}=(-0.5\sigma_V,0,0)$ and $\mathbf{r_2}=(0.5\sigma_V,0,0)$, we can calculate the spatial distribution function $g(x_3,y_3,z_3)$  around contact dimer. Fig.\ref{spmap} shows the spatial distribution in $xy$ plane calculated from Eq.(\ref{SA}). Here, $g(\mathbf{r_2},\mathbf{r_3})$ and $g(\mathbf{r_3},\mathbf{r_1})$ are obtained using the radial distribution functions $g_{VV}(r)$ caluculated by GCMC simulation. Compared to Fig.\ref{map}(a), the shape of density waves is in good agreement. This suggests that for simple solute models, the shape of the distribution of solvents can be estimated more or less accurately from the superposition approximation (SA).

Fig.\ref{gxy_SA} shows the spatial distributions calculated using the MHNC approximation and the GCMC simulation with the results obtained using Eq.(\ref{SA}), namely SA. The features of functions$g_{UV}(x,0,0)$ are similar. The locations for the peaks and the minimums are almost the same. The numerical agreement between the GCMC and MHNC results is excellent (the contact value $g(1.5\sigma,0,0)=3.8935$(SA) and $3.5062$(GCMC), $3.3838$(MHNC) in Fig.\ref{gxy},\ref{gxy_SA}). On the other hand, the deviation of the SA results from the exact results is largest. However, the deviation is smaller than those obtained using PY or HNC ($g(1.5\sigma,0,0)=3.0473$(PY) and $4.3583$(HNC) in Fig.\ref{gxy}).

We can find the same behavior as $g_{UV}(x,0,0)$ in $g_{UV}(0,y,0)$, and the differences become more significant. The first peaks of functions $g_{UV}(0,y,0)$ are higher than those of the $g_{UV}(x,0,0)$ because of the gain of the excluded volume at the contact position. Features of three functions$g_{UV}(0,y,0)$ for SA, MHNC, and GCMC are similar to each other again (the contact value $g(0,\sqrt{3}/2\sigma,0)=12.1041$(SA) and $8.7112$(GCMC), $7.1283$(MHNC) in Fig.\ref{gxy},\ref{gxy_SA}). The spatial distributions obtained by the SA with $g(r)$ obtained GCMC (or the MHNC approximation) are more accurate than those obtained using the HNC or the PY approximations ($g(0,\sqrt{3}/2\sigma,0)=4.8957$(PY) and $13.3187$(HNC) in Fig.\ref{gxy}). It means that the SA using the accurate $g(r)$ is not bad despite the simplicity when we discuss the qualitative behavior of the distribution functions. On the other hand, the deviations from the SA results show the existence of multiple body correlations .


Here, we discuss the approximations based on the three-body effect. The SA includes only three independent pair correlations and ignores the effect of any of these pair correlations. Thus, the SA differs from the exact result when the third particle is in the neighborhood of the pair. To assess the actual three-body effect, we consider the ratio $g^{(3)}$ to the value of the SA(Eq.(\ref{SA}))\cite{ueharab1979triplet, attard1992three, bildstein1994triplet}:

\begin{equation}
    \Gamma(\mathbf{r_1},\mathbf{r_2},\mathbf{r_3})=\frac{g^{(3)}(\mathbf{r_1},\mathbf{r_2},\mathbf{r_3})}{g(\mathbf{r_1},\mathbf{r_2})g(\mathbf{r_2},\mathbf{r_3})g(\mathbf{r_3},\mathbf{r_1})}
\end{equation}
which can be written as

\begin{equation}
    \Gamma(r_{12};\mathbf{r_3})=\frac{g^{(3)}(r_{12};\mathbf{r_3})}{g(\mathbf{r_2},\mathbf{r_3})g(\mathbf{r_3},\mathbf{r_1})}.
    \label{SAeq}
\end{equation}
If the SA were exact, the ratio $\Gamma$ should be unity. However, the calculated results are not unity except the SA.

\begin{figure}[h]
  \centering
  \includegraphics[width=4cm]{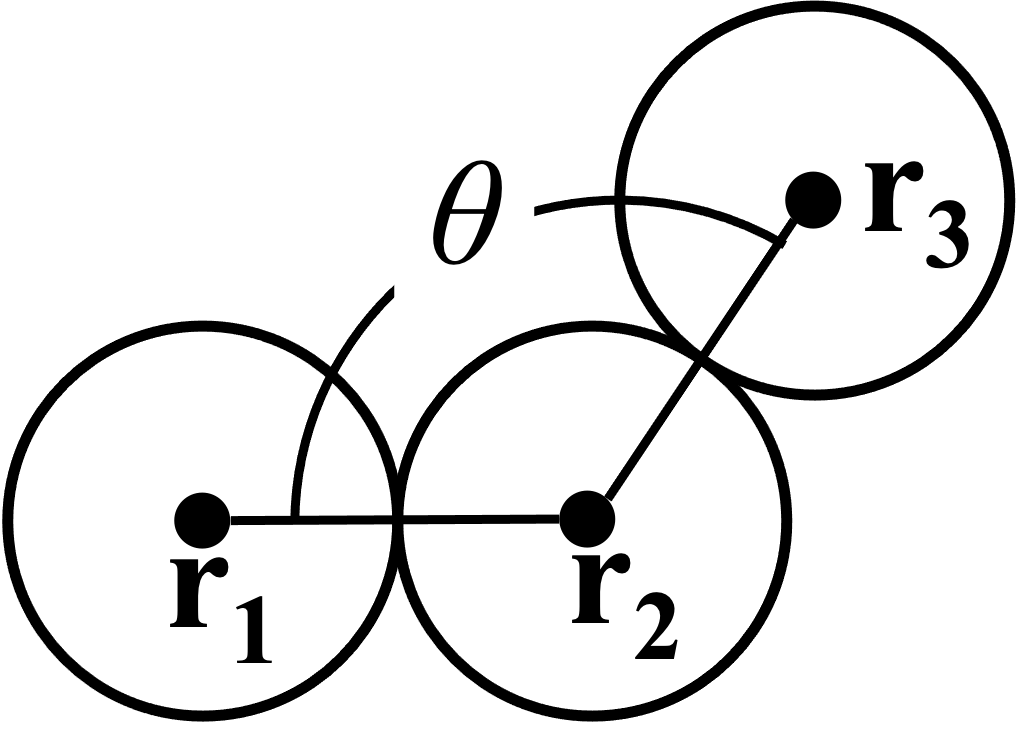}
  \caption{Discription of $\theta$}
  \label{diagram1}
\end{figure}

\begin{figure}[h]
  \begin{minipage}[t]{0.48\columnwidth}
    \centering
    \includegraphics[width=\columnwidth]
    {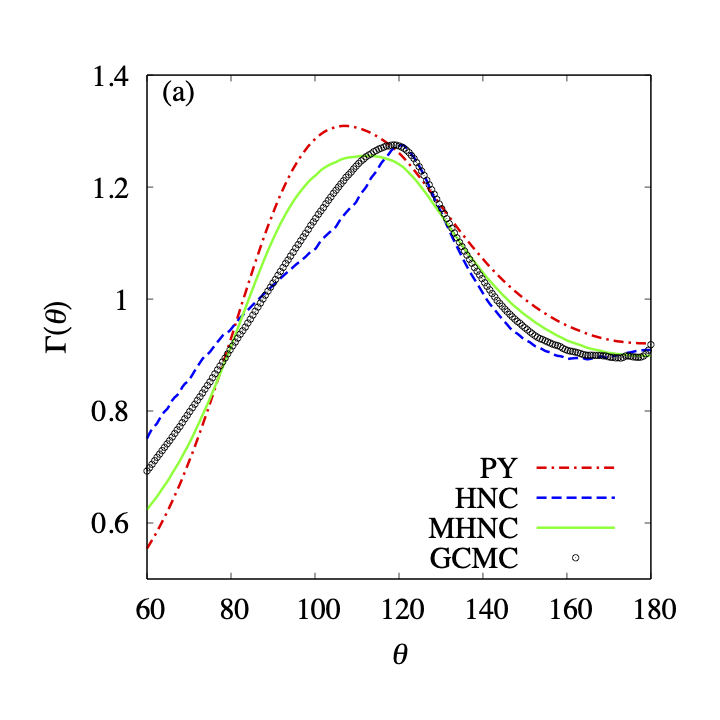}
  \end{minipage}
  \begin{minipage}[t]{0.48\columnwidth}
    \centering
    \includegraphics[width=\columnwidth]
    {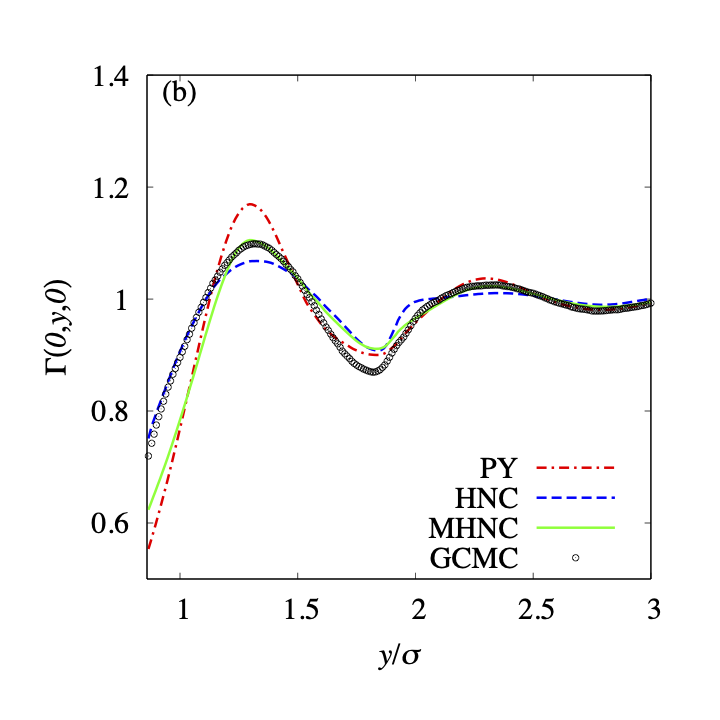}
  \end{minipage}
  \caption{The ratio (a) $\Gamma(\theta)$ and (b) $\Gamma(0,y,0)$ obtained by the GCMC method and the integral equation theory with the PY, HNC and MHNC closures.}
  \label{gamma}
\end{figure}

Here we think of a contact dimer (particles 1 and 2) and a solvent particle 3. We define the ratio $\Gamma(\theta)=\Gamma(\mathbf{r_3})$ as a function of the angle $\theta$ enclosed by $\mathbf{r_1}-\mathbf{r_2}$ and $\mathbf{r_3}-\mathbf{r_2}$ (see Fig.\ref{diagram1}). The calculated results $\Gamma(\theta)$ at $|\mathbf{r_3}-\mathbf{r_2}|=\sigma$ for the HNC, PY, and MHNC closures are shown in Fig.\ref{gamma}(a). The result of GCMC is also shown in the figure. First, we compare the SA and GCMC results. The exact result (GCMC) has two minimums around $60^\circ$ and $180^\circ$, and there is a peak around $120^\circ$. The first minimum at $60^\circ$ is caused by the overestimation of the SA. In other words, the numerator is smaller than the denominator in Eq.(\ref{SAeq}), which is the definition of $\Gamma$. Note the reference value, namely the denominator, is obtained by the SA in Eq.(\ref{SAeq}).  Actually, particle 3 is stable at $60^\circ$ because of the reduction of excluded volume for the three particles. However, the SA does not contain the exclusion effect for particle 3 by particle 2. Thus, the peak of the solvent spatial distribution estimated by the SA becomes larger than the exact value at 60°. Therefore, the minimum at $60^\circ$ appears in the $\Gamma(\theta)$\cite{footnote2}. On the other hand, it seems that the second minimum at $180^\circ$ appears because of the reductions of the three-body effect at the configuration because the value smoothly approaches unity.

The peak for the GCMC result around $120^\circ$ appears because of the effect of the solvent located at $60^\circ$. The reduced spatial distribution has a high peak at $60^\circ$, and the peak value is much larger than the unity. As the probability of the existence of the solvent particle at $60^\circ$ becomes larger, the fourth particle at $120^\circ$ becomes stabler because of the reduction of the excluded volume caused by the adsorption. The SA does not include this three-body effect. Therefore, the probability of particle existence obtained using the GCMC at $120^\circ$ becomes larger than that of the estimation by the SA, and the peak around $120^\circ$ in Fig.\ref{gamma}(a) appears.

Fig.\ref{gamma}(a) also has the plots $\Gamma(\theta)$ for various closures, namely the PY, the HNC, and the MHNC closures. The functions $\Gamma(\theta)$ maintain the same features: two minimums at $60^\circ$ and $180^\circ$ and one peak at $120^\circ$. Quantitative deviations can be seen in contrast to the qualitative validity of these approximate three-body correlations. The deviation of the PY result from the exact result(the GCMC simulation) is the largest. Although the sign of deviation for the HNC result is opposite to that for the MHNC result, both differences are small. The exact result is also close to these two plots. The HNC result is very accurate in the plots $\Gamma(\theta)$. However, it does not mean that the HNC closure is very accurate in the calculation of the spatial distribution. It should be noted that the deviation of the spatial distribution function calculated by the HNC closure is larger than that calculated by the MHNC closure due to the large deviation of the two-body correlation from the exact solution.  

Fig.\ref{gamma}(b) shows the ratio $\Gamma(\mathbf{r_3})=\Gamma(x,y,z)$ as a function of the coordinate $(x,y,z)$. $\Gamma(0,y,0)$ at $y=\sqrt{3}/2$ is equivalent to $\Gamma(\theta)$ at $\theta=60^\circ$. The plots $\Gamma(0,y,0)$ oscillate, and the qualitative features of the plots in the Fig.\ref{gamma}(b) are similar. Two deep minimums are located around $y=\sqrt 3/2$ and $1.8$, and a clear peak is located at $1.3$. However, the deviation of the PY result from the exact results (GCMC) is the largest. The accuracies of the MHNC and the HNC approximations depend on the value $y$. Around $y=\sqrt 3/2$ the HNC approximation is the most precise, while the MHNC becomes the most accurate around the first peak at $y= 1.8$. However, the difference is not quantitively large.

Here, we discuss the detail of the differences of $\Gamma(0,y,0)$. The MHNC results evaluated well the value of $\Gamma(0,y,0)$ near the first and second peaks. These peak positions correspond to the bottoms (or minimums) of the distribution function $g(0,y,0)$, where SA underestimates. We calculated the $g(r)$ between two hard spheres immersed in a hard-sphere fluid using the MHNC approximation in the previous study\cite{nakamura2019}. The MHNC results showed very accurate first minimums in the $g(r)$. The accuracies were much better than that of HNC results. We can conclude that these share common features. Therefore, the evaluation of the distribution functions using the MHNC approximation is very accurate, even if we discuss the value around the minimum. On the other hand, the deviation of the HNC results for $\Gamma(0,y,0)$ around the first minimum is smaller than that of the MHNC result, although the deviations of the HNC results are much larger than those of the MHNC results in comparing the functions $g(r)$. In the case of the HNC results, both the numerator and the denominator of Eq.(\ref{SAeq}) are overestimated. In contrast, only the denominator for the MHNC result is very accurate in the calculation Eq.(\ref{SAeq}).

\begin{figure}[h]
  \begin{minipage}[t]{0.48\columnwidth}
    \centering
    \includegraphics[width=\columnwidth]
    {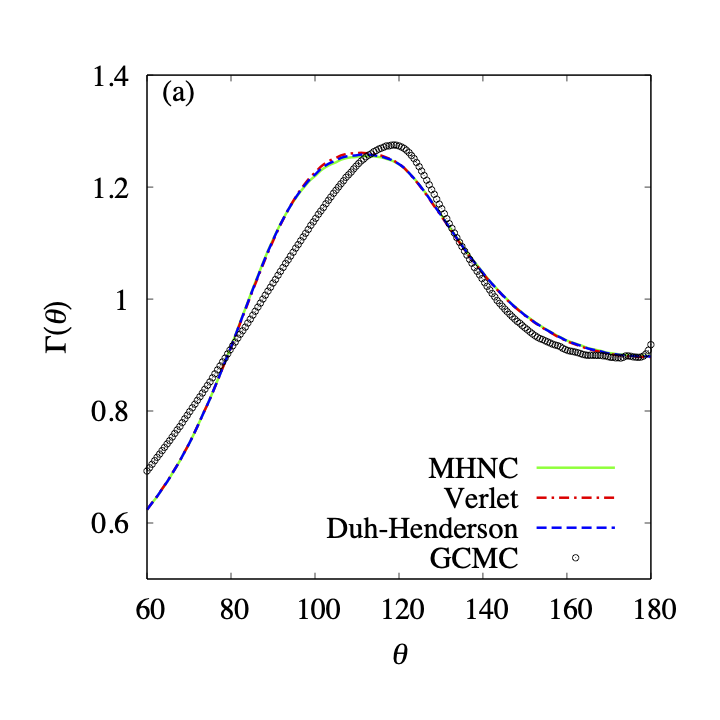}
  \end{minipage}
  \begin{minipage}[t]{0.48\columnwidth}
    \centering
    \includegraphics[width=\columnwidth]
    {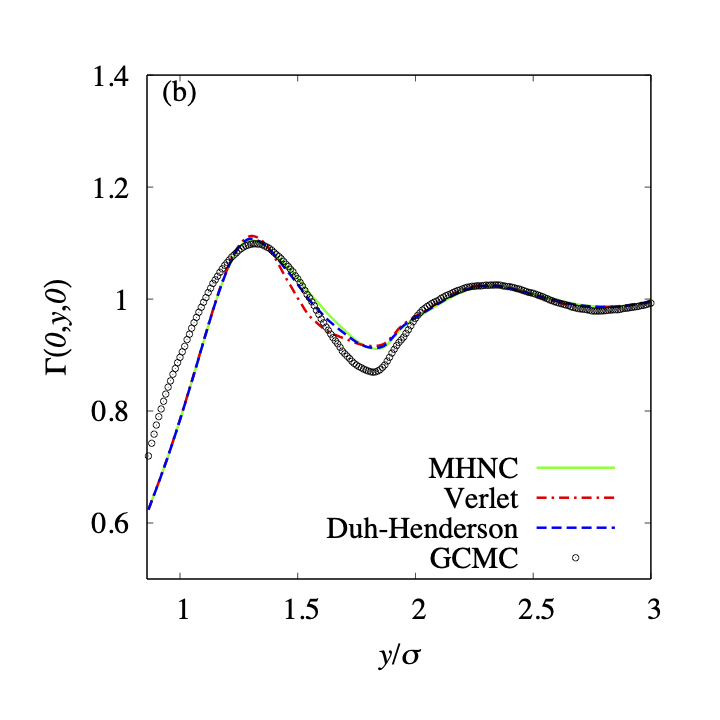}
  \end{minipage}
  \caption{The ratio $\Gamma(\theta)$ and $\Gamma(0,y,0)$ obtained by the GCMC method and the integral equation theory with the MHNC, Verlet and Duh-Henderson bridge functions.}
  \label{gamma_VD}
\end{figure}

Fig.\ref{gamma_VD} shows $\Gamma(\theta)$ and $\Gamma(0,y,0)$ of the GCMC result and three results for the various closures with bridge functions: the MHNC, the Verlet, and the Duh-Henderson bridge. The three closures have virtually the same results. We can find differences in $\Gamma(0,y,0)$ around $y=1.5\sigma_V$. The deviation of the Verlet bridge result from the exact result is the largest. However, the difference can be ignorable. Therefore, it is consistent with the agreement in the spatial distribution functions (see Fig.\ref{gxy-VD}).

It seems that the HNC approximation overestimates the pair correlation function due to the ignorance of the bridge function (see Eq. (\ref{HNCeq}).) On the other hand, it has been shown that the pair correlation function given by the MHNC approximation is very accurate in the present system\cite{nakamura2019} . Furthermore, it is known that the pair correlation function also agrees well with MC if the bridge function was constructed to satisfy the thermodynamic consistency\cite{rogers1984, hansen1985}. The HNC approximation is also insufficient for thermodynamic consistency, but the results given by the MHNC approximation are automatically almost completely satisfied \cite{hayashi2016}. Therefore, we expected that the bridge function of the MHNC approximation is an adequate improvement and that the three-body correlation is also better than that calculated using the HNC approximation.

However, the superiority of some closures with a bridge function in the spatial distribution functions is not caused by the superiority of the three-body correlations. Here, the validity of the closure can also be evaluated in terms of the three-body correlation. Though the incorporation of the Verlet, Duh-Henderson, and Kinoshita bridge functions improves the pair correlation function, it does not necessarily lead to better results for the triplet distribution function. It is interesting to compare the results from the MC simulation and the OZ equations coupled with PY, HNC, and the three different closures in terms of the triplet distribution function. Under the present calculation conditions, the calculated results do not always mean that the MHNC approximation gives better results than the HNC approximation for the three-body correlation.
\begin{figure}[h]
  \begin{minipage}[t]{0.48\columnwidth}
    \includegraphics[width=0.9\columnwidth]
    {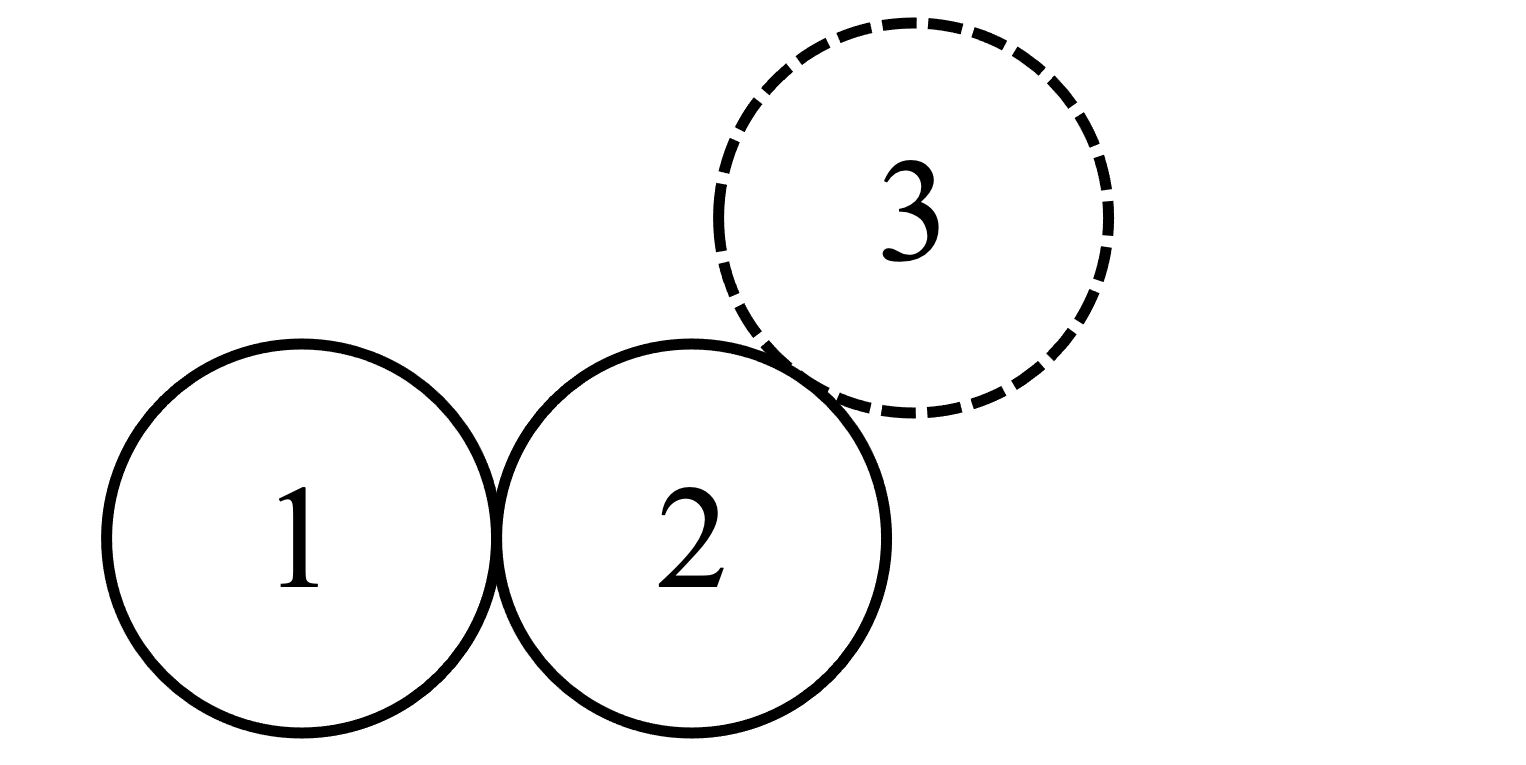}
    \begin{flushleft}
    (a) Configuration 1 \\
    : Particle 1 and 2 are fixed
    \end{flushleft}
  \end{minipage}
  \begin{minipage}[t]{0.48\columnwidth}
    \centering
    \includegraphics[width=0.9\columnwidth]
    {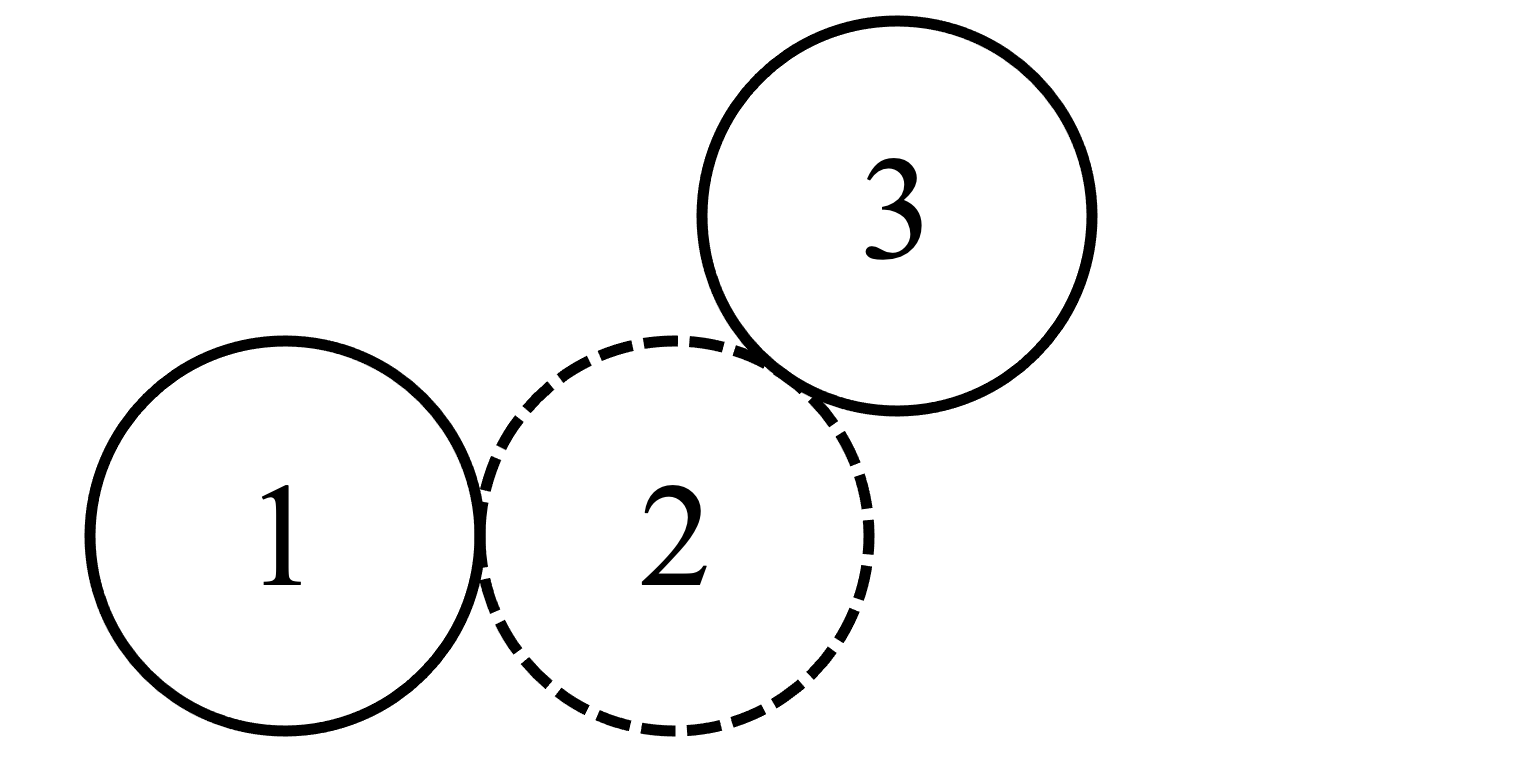}
    \begin{flushleft}
    (b) Configuration 2 \\
    : Particle 1 and 3 are fixed
    \end{flushleft}
  \end{minipage}
  \caption{Discription of the two configuration for calculating $\Gamma(\theta)$: (a)$\Gamma(\theta)=\Gamma(r_{12};\mathbf{r_3})$ and (b)$\Gamma(\theta)=\Gamma(r_{13};\mathbf{r_2})$. The solid spheres are fixed and the broken spheres are obtained for distribution functions.}
  \label{diagram2}
\end{figure}

\begin{figure}[h]
    \centering
    \includegraphics[width=7.5cm]{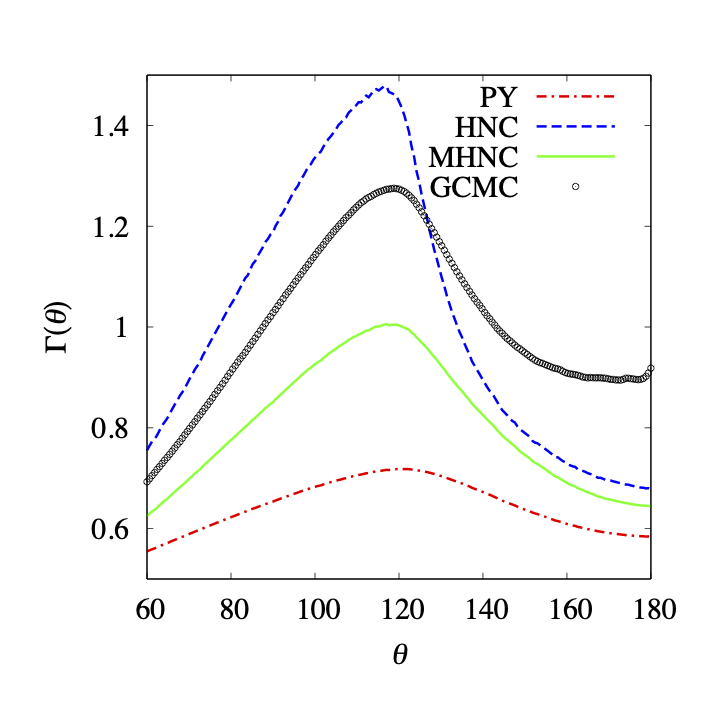}
    \caption{The ratio $\Gamma(\theta)$ calculated by the integral equation theory with the PY, HNC, and MHNC closures at configuration 2 and the GCMC method.}
    \label{gammare}
\end{figure}

In the above calculation, configuration 1 (see Fig.\ref{diagram2}(a)) was adopted. We can also obtain the $\Gamma(\theta)$ by using configuration 2 (see Fig\ref{diagram2}(b)). 
In the method, a pair of two separate particles 1 and 3, fixed at ${\rm r}_{1}$ and ${\rm r}_{3}$ is immersed in the solvent particles. Thus, we calculate the spatial distribution function around them $g^{(3)}(r_{13}; {\rm r}_{2})$ to obtain $\Gamma (\theta)$. In other word, $\Gamma(\theta)$ is calculated using the spatial distribution function at $\mathbf{r}_2$ when particle 1 and 3 is located at variable $\theta$. Then, $\Gamma(\theta)$ is written as

\begin{equation}
    \Gamma(\theta)=\Gamma(r_{13};\mathbf{r_2})=\frac{g^{(3)}(r_{13};\mathbf{r_2})}{g(\mathbf{r_1},\mathbf{r_2})g(\mathbf{r_2},\mathbf{r_3})}.
\end{equation}
Here,  $g^{(3)}(r_{13}; r_{2})$ is not the spatial distribution around a contact dimer solute. However, we can compare the $\Gamma (\theta)$ for configuration 2 with that for configuration 1. In Fig.\ref{gammare}, we show the $\Gamma(\theta)$ results when the configuration 2 are adopted. The discrepancies between the exact GCMC result and the results for the various closures are much larger than those obtained using configuration 1 (Fig.\ref{gamma}).

Here, we focus on the difference between the calculated results using the integral theories and the exact result using GCMC in Fig.\ref{gammare}. In the case of exact results, the plot for configuration 2 is the same as that for configuration 1. In the case of configuration 1, we can also find the closure dependence (See Fig.\ref{gamma}(a)). However, the dependence is much smaller than that for configuration 2 (See Fig.\ref{gammare}). In the case of configuration 2, the value calculated using the MHNC is about $0.65$ when $\theta=180^\circ$. At $\theta=180^\circ$, since the three particles are aligned on a straight line, we expect that the three-body correlation effect is less apparent. It is because particle 3 is not strongly affected by particle 1 since particle 3 is located just behind particle 2 (See Fig.\ref{diagram2}). In fact, in the case of configuration 1, all plots converge to a value of about $0.9$ at $\theta=180^\circ$ (See Fig.\ref{gamma}(a)). That is, in the case of configuration 1, especially when $\theta=180^\circ$, any approximation is close to the exact result. These results contrast with that of configuration 2. The poor approximation for the three-body correlation by closures is emphasized using configuration 2.

We discuss the reason for the difference between configurations 1 and 2. In a dilute system, the three-body correlation becomes weak, and $\Gamma$ goes to 1\cite{kubota2012}. Conversely, the multi-body correlation becomes important in a dense fluid. Then, the value $\Gamma$ is expected to worsen as the local density becomes larger. As the overlap of the excluded volume increase, the configuration of the three particles becomes stable in the present packing fraction, and the local density of the third particle becomes high at the location. Therefore, the amount of the excluded volume overlap for the third particle, namely particle 2 for configuration 2, should correlate with the accuracy of the multi-body correlation.

For example, in the case of configuration 2, particle 2 contacts particles 1 and 3. Then, particle 2 has two excluded volume overlaps with particles 1 and 3 at any angle. Because of the large overlap amount, the local density of the third particle for configuration 2 is high, and the three-body correlation value deviates from the exact value in the entire range of $\theta=60^\circ$ to $180^\circ$ except the cross point near $\theta=130^\circ$ for HNC closure.

Let us discuss the validity of the above argument regarding the relationship between local density and three-body correlations based on the results for configuration 1. In the configuration, the excluded volume of particle 3 overlaps with both those of particles 1 and 2 near $\theta=60^\circ$. Therefore, at $60^\circ<\theta<120^\circ$, we can find differences between the exact calculation results and those calculated using the integral equation theories. On the other hand, when $120^\circ<\theta<180^\circ$, the excluded volume for particle 3 overlaps only that of particle 2. Since the overlap is less than that for $60^\circ<\theta<120^\circ$, the results of the integral equation theory roughly agree with the exact results. The above results coincide with the argument that the three-body correlation tends to be inaccurate in configurations where the local density of particles is high.

\section{Conclusion}
We calculated the reduced spatial density profile of hard spheres near a hard-sphere dimer using some closures. Using some closures, we calculated the reduced spatial density profile of hard spheres near a hard-sphere dimer. We examined not only the PY and the HNC closure but also the closures with bridge functions proposed by Verlet, Duh-Henderson, and Kinoshita. The solute-solvent spatial density profiles are obtained three-dimensional OZ equation coupled with those closures because the hard-sphere dimer solute is not spherical. The results were compared with the reduced density profile calculated using the GCMC simulation.

The comparison with the exact result obtained using the GCMC indicates that the three closures taking account of bridge function, such as the MHNC closure, are much better than others. The  SA is inferior to the three closures taking account of the bridge function in terms of accuracy. Because, the calculation cost of the SA is not expensive,  it is useful in calculating the spatial distribution function. If we use the precise radial distribution function $g(r)$, the reduced density profile obtained by the SA with the function $g(r)$ is better than those obtained using two traditional closure relations, namely the PY and HNC closures.  However, the closures with bridge functions have an advantage in the calculation of the reduced density profiles $g_{\rm UV} (x,y,z)$. As we discussed above, the advantage of the bridge function proposed by Kinoshita should appear under other conditions.

The three-body correlations are also compared with the exact results. The SA and the PY approximation are inferior to the NHC and other closures. However, we cannot find the superiority of the three closures taking into account the bridge function to the results using the HNC closure. Therefore, it seems that the HNC approximation overestimates the pair correlation function due to the ignorance of the bridge function. On the other hand, under the present calculation conditions, the closures with the bridge functions, such as the MHNC approximation, seems slightly worse than the HNC approximation for the three-body correlation, although the results given by the MHNC approximation are almost completely accurate. Because the HNC approximation is insufficient for thermodynamic consistency, the slight worsening caused by the bridge function in the MHNC approximation is surprising. It seems that there is a cancellation due to the deviations on the denominator and the numerator of Eq. (\ref{SAeq}) in the case of the HNC approximation.

\bibliography{ref}

\begin{thebibliography}{10}
\bibitem{footnote1}
This effective interaction becomes purely enthalpic under isobaric condition. However, it does not mean that the effective interaction is driven by energy. The translational motion of depletants is the origin, even under the isobaric condition.
\bibitem{footnote2}
This minimum can also be explained in terms of the triplet overlap of the excluded volume. In the SA, the overlap of the excluded volume is double-counted.
\end{thebibliography}
\bibliographystyle{jpsj}

\end{document}